\documentclass[12pt]{article}

\usepackage{amsfonts,amscd,amssymb,amsmath,latexsym}
\usepackage{verbatim}

\newcommand{\Real}{\mathbb{R}}
\newcommand{\dx}{d^4x}

\newcommand{\pmu}{\partial_{\mu}}
\newcommand{\pnu}{\partial_{\nu}}
\newcommand{\nmu}{\nabla_{\mu}}
\newcommand{\nnu}{\nabla_{\nu}}
\newcommand{\FF}{\mathcal{F}}

\usepackage{graphicx}
\newcommand{\miktex}{\hbox{Mik\kern-.15em\TeX}}

\providecommand{\rjpurl}[1]{\bgroup\footnotesize\ttfamily #1\egroup}

\title{New Cosmological Solutions in \\ Nonlocal Modified Gravity } 

\author{{I. Dimitrijevic$^a$, B. Dragovich$^b$, J. Grujic$^c$ and Z. Rakic$^a$} \\  \\
$^a$Faculty of Mathematics, University of Belgrade \\
Studentski trg 16, Belgrade, Serbia\\
{\em Email}: ivand@matf.bg.ac.rs \\
$^b$Institute of Physics, University of Belgrade \\ Pregrevica 118, 11080 Zemun, Belgrade, Serbia\\
{\em Email}: dragovich@ipb.ac.rs\\
$^c$Teachers Training Faculty, University of Belgrade \\ Kraljice Natalije 43, Belgrade, Serbia\\
{\em Email}: jelenagg@gmail.com}

\date{}

\begin{document}
\maketitle

\begin{abstract}
We consider  some cosmological aspects of nonlocal modified gravity with $\Lambda$ term,
where nonlocality is of the type $R \mathcal{F}(\Box) R$. Using ansatz of the form $\Box R = r R +s,$ we
find a few $a(t)$ nonsingular bounce cosmological solutions for all three values of spatial curvature parameter $k$.
We also discuss this modified gravity model from $F(R)$ theory point of view.
\end{abstract}

\section{Introduction}

Modern theory of gravity is general theory of relativity, which was founded by Einstein at the end of
1915 and has been successfully confirmed for the Solar System. It is given by the Einstein equations
of motion for gravitational field  $g_{\mu\nu}$:
$
R_{\mu\nu} - \frac{1}{2} R g_{\mu\nu} = 8 \pi G T_{\mu\nu},
$
which can be derived from the Einstein-Hilbert action $S =
\frac{1}{16 \pi G} \int \sqrt{-g} R d^4x + \int \sqrt{-g}
\mathcal{L}_{mat} d^4x , $ where $g = det (g_{\mu\nu})$ and $c = 1$.

Attempts to modify  Einstein's theory of gravity started already at its very
beginning  and it was mainly motivated by  investigation of  possible
mathematical generalizations. During last decade  there has been an intensive
activity in gravity modification, motivated by discovery of
accelerating expansion of the Universe, which has not yet
generally accepted theoretical explanation. If Einstein's  gravity
is theory of gravity for the Universe as a whole then it has to be
some new kind of matter with negative pressure, called {\it dark
energy}, which is responsible for the accelerated  Universe expansion.
However, general
relativity has not been verified at the cosmic scale (low
curvature regime) and dark energy has not been confirmed in a laboratory.
Hence, after discovery of the accelerated Universe there emerged  a renewed  interest in modification of
Einstein's theory of gravity, which should be some kind of its
generalization (for a recent review of various approaches, see
\cite{clifton}). However there is not a unique way how to modify
the Einstein-Hilbert action. Among many approaches there are two of them,
which have attracted more interest than the others: 1) $F(R)$ theories of gravity
(for a review, see \cite{faraoni,nojiri}) and 2) nonlocal gravities (see,
e.g. \cite{nojiri,biswas0,biswas,koshelev1,koshelev2,biswas1,modesto} and references therein).

In the case of $F(R)$ gravity, the Ricci scalar $R$ in the action
is replaced by a function $F(R)$. This has been extensively investigated
for the various forms of function $F(R)$.

\label{sec:1}

In the sequel we shall consider some cosmological aspects of a nonlocal gravity.
Here, nonlocality means that Lagrangian contains an infinite number of
space-time derivatives, i.e. derivatives up  to an infinitive
order in the form of d'Alembert operator $\Box$ which is argument of an analytic function.
In string theory
nonlocality emerges as a consequence of extendedness of strings.
Since string theory contains gravity, as well as other kinds of
interaction and matter, it is natural to expect nonlocality not
only in the matter sector   but also in  geometrical sector of
gravity. So the main motivation to consider an additional nonlocal term in the
Einstein-Hilbert action comes from the string theory. On some developments in cosmology with nonlocality in
the matter sector one can see, e.g.,
\cite{arefeva,calcagni,barnaby,koshelev,calcagni1} and references therein.
In the next sections we shall discuss a nonlocal modification of
only geometry sector of gravity and its corresponding new cosmological
solutions.

\section{A Nonlocal Modification of Gravity}

\label{sec:2}

In our consideration, nonlocal modification of gravity is a replacement
of the Ricci curvature $R$ in the Einstein-Hilbert action  by a suitable function
$F (R, \Box), $ where $\Box = \frac{1}{\sqrt{-g}} \partial_{\mu}
\sqrt{-g} g^{\mu\nu} \partial_{\nu} \, $ is d'Alembert-Beltrami operator.

In this paper we consider nonlocal gravity model
without matter, given by the action in the form
\begin{equation} \label{lag:1}
S =  \int d^{4}x \sqrt{-g}\Big(\frac{R - 2 \Lambda}{16 \pi G} +
\frac{C}{2}R \mathcal{F}(\Box)R  \Big),
\end{equation}
 where  $ \mathcal{F}(\Box)=
\displaystyle \sum_{n =0}^{\infty} f_{n}\Box^{n}$  is an analytic function  of the d'Alembert-Beltrami operator and $C$ is a
constant.
Study of this model \eqref{lag:1} was proposed in \cite{biswas0} and some further   developments are presented in
\cite{biswas,koshelev1,koshelev2,biswas1}.
This model is attractive because it is ghost free and has some nonsingular bounce solutions, which can solve
the Big Bang cosmological singularity problem.

 By variation of the action (\ref{lag:1}) with
respect to metric $g_{\mu\nu}$ one obtains the corresponding equation of motion:

\begin{equation} \begin{aligned} \label{eom:1}
&C \Big( 2 R_{\mu\nu} \mathcal{F}(\Box) R - 2(\nmu\nnu -
g_{\mu\nu} \Box)( \mathcal{F}(\Box) R) - \frac{1}{2}
g_{\mu\nu} R \mathcal{F}(\Box) R  \\
&+ \sum_{n=1}^{\infty} \frac{f_n}{2} \sum_{l=0}^{n-1} \big(
g_{\mu\nu} \left( g^{\alpha\beta}\partial_{\alpha} \Box^l R
\partial_{\beta} \Box^{n-1-l} R + \Box^l R \Box^{n-l} R
\right) \\
&- 2 \pmu \Box^l R \pnu \Box^{n-1-l} R\big) \Big) = \frac{-1}{8
\pi G} (G_{\mu\nu} + \Lambda g_{\mu\nu}).
\end{aligned} \end{equation}

The trace  and $00$ component of \eqref{eom:1} are:
\begin{equation} \label{trace:1}
6\Box ( \mathcal{F}(\Box) R) + \sum_{n=1}^{\infty} f_n
\sum_{l=0}^{n-1} \left(
\partial_{\mu} \Box^l R
\partial^{\mu} \Box^{n-1-l} R + 2 \Box^l R \Box^{n-l} R
\right)  = \frac{1}{8 \pi G C} R - \frac{\Lambda}{2 \pi G C},
\end{equation}
\begin{equation} \begin{aligned} \label{eom:00}
&C \Big( 2 R_{00} \mathcal{F}(\Box) R - 2(\nabla_0
\nabla_0 - g_{00} \Box)( \mathcal{F}(\Box) R) - \frac{1}{2}
g_{00} R \mathcal{F}(\Box) R  \\
&+ \sum_{n=1}^{\infty} \frac{f_n}{2} \sum_{l=0}^{n-1} \big( g_{00}
\left( g^{\alpha\beta}\partial_{\alpha} \Box^l R
\partial_{\beta} \Box^{n-1-l} R + \Box^l R \Box^{n-l} R
\right)  \\
&- 2 \partial_0 \Box^l R \partial_0 \Box^{n-1-l} R\big) \Big) =
\frac{-1}{8 \pi G}( G_{00} + \Lambda g_{00}).
\end{aligned} \end{equation}

We use Friedmann-Lema\^{\i}tre-Robertson-Walker (FLRW) metric
$ds^2 = - dt^2 + a^2(t)\big(\frac{dr^2}{1-k r^2} + r^2 d\theta^2 +
r^2 \sin^2 \theta d\phi^2\big)$ and investigate all three
possibilities for curvature parameter $k$ ($0,\pm 1$).


\section{Ansatz and  Solutions}

Investigation of equation \eqref{eom:1} and finding its  solutions is
a very difficult task. In the case of the FLRW flat metric $(k=0)$ two nonsingular cosmological solutions
for the scale factor are found: $a(t) = a_0 \cosh{\left(\sqrt\frac{\Lambda}{3}t\right)}$, see \cite{biswas0,biswas},
and $a(t) = a_0 e^{\frac{1}{2}\sqrt{\frac{\Lambda}{3}}t^2}$, see \cite{koshelev2}.

To get some new solutions we also use ansatz of the form

\begin{equation}  \label{ansatz}\Box R = r R + s, \end{equation}
proposed in \cite{biswas}, where $r$ and $s$ are real parameters that will
be fixed later. The first two consequences of this ansatz are

\begin{equation}\label{nth degree}
\Box^{n} R = r^{n}(R +\frac sr ) , \, n\geq 1 , \,  \qquad \, \mathcal{F}(\Box)
R = \mathcal{F}(r) R + \frac sr(\FF(r)-f_0) .
\end{equation}

Now we can search for a solution of the scale factor $a(t)$ in the form of a linear combination
of $e^{\lambda t}$ and $e^{-\lambda t}$,  i.e.
\begin{equation} \label{sol:a}
a(t) = a_0  (\sigma  e^{\lambda t} + \tau e^{-\lambda t} ), \quad
0< a_0, \lambda,\sigma,\tau \in \Real .
\end{equation}
Then  the corresponding expressions for the Hubble parameter $H(t) = \frac{\dot{a}}{a},$ scalar curvature
$R(t) = \frac{6}{a^2} (a \ddot{a} + \dot{a}^2 + k) $ and $\Box R$ are:
\begin{equation} \begin{aligned} \label{sol:all}
H(t) &= \frac{\lambda  (\sigma  e^{\lambda t} - \tau e^{-
\lambda t}) } {\sigma  e^{\lambda t} + \tau e^{- \lambda t}}, \\
R(t) &= \frac{6 \left(2 a_0^2 \lambda ^2 \left(\sigma^2 e^{4 t
\lambda }+\tau ^2\right)+k e^{2 t \lambda }\right)}{a_0^2 \left(\sigma
e^{2 t \lambda }+\tau \right)^2},\\
\Box R &= -\frac{12 \lambda ^2 e^{2 t \lambda } \left(4 a_0^2
\lambda ^2 \sigma  \tau -k\right)}{a_0^2 \left(\sigma  e^{2 t \lambda
}+\tau \right)^2}.
\end{aligned} \end{equation}
We can rewrite $\Box R$ as
\begin{equation}\label{ansatz:1}
\Box R = 2\lambda^2 R - 24\lambda ^ 4 , \qquad r = 2\lambda^2 , \, \, s = - 24\lambda ^ 4.
\end{equation}

Substituting parameters $r$ and $s$  from \eqref{ansatz:1} into
\eqref{nth degree} we obtain

\begin{equation} \begin{aligned} \label{sol-all}
\Box^n R &= (2\lambda^2)^n (R - 12\lambda ^2) , \, \, n \geq 1 ,  \\
\mathcal{F}(\Box) R &= \mathcal{F}(2 \lambda^2)R - 12
\lambda^2(\mathcal{F}(2 \lambda^2) - f_0).
\end{aligned} \end{equation}

Using this in \eqref{trace:1} and \eqref{eom:00} we obtain
\begin{align} \label{trace:2}
&36\lambda^2 \mathcal{F}(2 \lambda^2) (R - 12\lambda ^2) +
\mathcal{F}'(2 \lambda^2) \left( 4 \lambda^2 (R - 12\lambda ^2)^2
- \dot R^2 \right)  \nonumber \\
&-24 \lambda^2 f_0(R - 12 \lambda^2) = \frac{R - 4\Lambda}{8 \pi G C} ,
\end{align}
\begin{align} \label{eom:2}
& (2 R_{00} + \frac{1}{2} R)\left( \mathcal{F}(2 \lambda^2)R - 12
\lambda^2(\mathcal{F}(2 \lambda^2) - f_0) \right) -\frac{1}{2}\mathcal{F}' (2 \lambda^2) \left( \dot R^2 + 2 \lambda^2 (R - 12 \lambda^2)^2 \right)  \nonumber \\
&-6\lambda^2(\FF(2\lambda^2)-f_0) (R-12\lambda^2) +6 H \FF(2 \lambda^2) \dot R= - \frac{1}{8\pi G C}( G_{00} - \Lambda) .
\end{align}

Substituting $a(t)$ from \eqref{sol:a} into equations \eqref{trace:2} and \eqref{eom:2} one obtains respectively the following two equations  as polynomials in $e^{2\lambda t}$:
\begin{align}
& \frac{a_0^4 \tau^6}{4\pi G} \left(3 \lambda^2 - \Lambda \right) + 3 a_0^2 \tau^4 Q_1 e^{2 \lambda t} + 6 a_0^2 \sigma
\tau^3 Q_2 e^{4\lambda t} -2 \sigma  \tau  Q_3 e^{6\lambda t}  \nonumber\\
&+ 6 a_0^2 \sigma^3 \tau Q_2 e^{8\lambda t} + 3 a_0^2 \sigma^4 Q_1 e^{10 \lambda t} +  \frac{a_0^4 \sigma^6}{4\pi G} \left(3 \lambda^2 - \Lambda \right) e^{12 \lambda t} = 0 \label{eq12} ,\\
&\frac{\tau^6 a_0^4}{8\pi G} \left(3 \lambda ^2- \Lambda \right) + 3 \tau^4 a_0^2 R_1 e^{2\lambda t} + 3 \tau^2 R_2 e^{4\lambda t} +
2\sigma\tau R_3 e^{6 \lambda t} \nonumber \\
&+ 3 \sigma^2 R_2 e^{8\lambda t} + 3 \sigma^4 a_0^2 R_1 e^{10 \lambda t} + \frac{\sigma^6 a_0^4}{8\pi G} \left(3 \lambda ^2- \Lambda \right) e^{12 \lambda t} = 0, \label{eq13}
\end{align}
where
\begin{align}
Q_1&= 36 C \lambda^2 K \FF(2\lambda^2)+ a_0^2(-96  C f_0 \lambda ^4 +\frac{\lambda^2}{\pi G}  -\frac {\Lambda }{2\pi G}) \sigma  \tau \nonumber\\
&+24 C f_0 k \lambda ^2+ \frac{k}{8\pi G} , \\
Q_2 &= 72 C \lambda^2 K \FF(2\lambda^2) + a_0^2(-192  C f_0 \lambda ^4 + \frac {7\lambda ^2}{8\pi G} -
\frac{5\Lambda}{8\pi G} ) \sigma  \tau \nonumber \\
&+48 C f_0 k \lambda ^2+ \frac{k}{4\pi G} , \\
Q_3 &= -324C a_0^2 \lambda^2 \sigma \tau K  \FF(2\lambda^2) + 144 C \lambda^2 K^2 \FF'(2\lambda^2)  \nonumber\\
&- a_0^2 k (216C f_0 \lambda^2 + \frac{9}{8\pi G}) \sigma \tau + a_0^4 (864 C f_0 \lambda^4 - \frac{3 \lambda^2}{\pi G} + \frac{5 \Lambda}{2\pi G}) \sigma^2 \tau^2 ,
\end{align}
\begin{align}
R_1 &= Q_1 - \frac{3\lambda^2 - \Lambda}{4\pi G} \sigma \tau a_0^2 ,\\
R_2&= -6 C \left(k-12 a_0^2 \lambda ^2 \sigma  \tau \right) K \FF(2\lambda^2) - 36 C \lambda ^2 K^2 \FF'(2\lambda^2)  \nonumber\\
& + \frac{a_0^2 k}{2 \pi  G} \left(192 \pi G C f_0 \lambda ^2 + 1\right) \sigma \tau - \frac{a_0^4}{8 \pi  G} \left(3072 \pi G C f_0 \lambda ^4+\lambda ^2+5 \Lambda \right) \sigma^2 \tau^2  ,\\
R_3&= - 18 C \left(k-6 a_0^2 \lambda ^2 \sigma  \tau \right) K \FF(2\lambda^2) + 36 C \lambda^2 K^2 \FF'(2\lambda^2)  \nonumber\\
&+ \frac{9 a_0^2 k}{8 \pi  G} \left(192 \pi G C f_0 \lambda ^2+1\right) \sigma \tau -\frac{a_0^4}{4 \pi  G} \left(3456 \pi G C f_0 \lambda ^4+3 \lambda ^2+5 \Lambda \right) \sigma^2 \tau^2,
\end{align}
and $K = 4 a_0^2 \lambda ^2 \sigma \tau  - k .$

Equations \eqref{eq12} and \eqref{eq13} are satisfied when $\lambda = \pm \sqrt{ \frac{\Lambda}{3}},$  as well as $Q_1 = Q_2 =Q_3 =0$ and $R_1 = R_2 =R_3 = 0.$
Note that this approach to find conditions under which solution exists differs with respect to approach used in \cite{biswas} and \cite{koshelev2}.


The corresponding solutions can  split into the following three cases.

\textit{Case} 1.
\begin{equation}  \label{sol:case1}
\FF\left( 2 \lambda^2 \right) = 0, \quad
\FF'\left(  2 \lambda^2 \right) = 0, \quad
f_0 = - \frac{1}{64 \pi G C\Lambda} .
 \end{equation}

\textit{Case} 2.
\begin{equation}\label{sol:case2}
3  k  = 4 a_0^2 \Lambda \sigma \tau .
 \end{equation}

\textit{Case} 3.
\begin{equation}  \label{sol:case3}
\FF\left(  2 \lambda^2 \right) = \frac{1}{96 \pi G C \Lambda} + \frac{2}{3} f_0, \quad
\FF'\left(  2 \lambda^2 \right) = 0, \quad
k  = - 4 a_0^2 \Lambda \sigma \tau . \end{equation}


In the first case we have  family of solutions for
arbitrary $\sigma, \tau $ and $a_0$
\begin{equation} \nonumber
a(t) = a_0 (\sigma e^{\lambda t} + \tau e^{- \lambda t})
\end{equation}
with function $\FF$  satisfying conditions given in
\eqref{sol:case1} and arbitrary $k = 0, \pm 1 .$ Thus it includes also
solution $a(t) = a_0 \cosh{\left(\sqrt\frac{\Lambda}{3}t\right)},$ which was found
in \cite{biswas0}  with addition of some radiation \cite{biswas} in the action.

The second case yields a family of solutions for arbitrary $\sigma
\neq 0$ and $a_0$
\begin{align*}
a(t) &= a_0 \left(\sigma e^{\lambda t} + \frac{3 k }{4a_0^2
\Lambda \sigma} e^{- \lambda t}\right)\\
\end{align*}
 which are valid for arbitrary analytic function $\FF$.

The third case yields another family of solutions
\begin{align*}
a(t) &= a_0 \left(\sigma e^{\lambda t} - \frac{ k }{4a_0^2
\Lambda \sigma} e^{- \lambda t}\right)\\
\end{align*}
and function $\FF$ has to satisfy conditions given in
\eqref{sol:case3}.

Note that if $k=0$ then equation \eqref{sol:case2} and third equation
in \eqref{sol:case3} coincide, $\sigma$ or $\tau$ has to be zero, and we have 2 solutions
\begin{equation} \nonumber
a_1(t) = a_0 e^{\lambda t} ,  \qquad
a_2(t) = a_0 e^{-\lambda t} ,
\end{equation}
where $\sigma$ is absorbed into $a_0$. These are the  de Sitter solutions, see also \cite{biswas1}.



\section{$F(R)$ Theory}

 Action
\eqref{lag:1} with ansatz \eqref{ansatz}  gives an  $F(R)$ theory and we get
another way to analyze the above solutions .
The  corresponding  action for $F(R)$ is
\begin{equation}
S' = \int \frac{\sqrt{-g}}{16 \pi G} F(R) \dx , \qquad F(R) = \alpha R^2 + \beta R - 2  \Lambda,  \label{lag:2}
\end{equation}
\begin{equation}
\alpha = 8 \pi G C \FF(2\lambda^2), \quad
\beta = 1 - 96 \pi G C \lambda^2 (\FF(2\lambda^2) - f_0) .\label{beta}
\end{equation}

By variation of the action \eqref{lag:2} we get the following
equation \cite{faraoni}:

\begin{equation}\label{fieldeq}
F'(R) R_{\mu\nu} - \frac{1}{2}F(R)g_{\mu\nu} - [ \nmu\nnu -
g_{\mu\nu}\Box ] F'(R) = 8 \pi G T_{\mu\nu} .
\end{equation}

The generalized Friedmann equations are
\begin{align*}
\big(\frac{\dot a}{a}\big)^{2} + \frac{k}{a^{2}} = \frac{8 \pi G}{3
F'(R)}
(\rho + \overline{\rho}), \\
\frac{\ddot a}{a} =  -\frac{ 4 \pi G}{3F'(R)} (\rho + 3p + \overline{\rho} +
3\overline{p}).
\end{align*}


From the equations of motion  we derive expressions
for effective density $\overline \rho$ and pressure $\overline p$
\begin{equation} \begin{aligned}
\overline \rho &= \frac{-1}{8 \pi G} \Big( \frac 12 F(R) - \frac
R2 F'(R) + 3 \frac{\dot a}{a} \dot R F''(R)\Big) , \\
\overline p &= \frac{1}{8 \pi G} \Big( \frac 12 F(R) - \frac R2
F'(R) - (3 \frac{\dot a}{a} \dot R + \ddot R) F''(R) - \dot R^2
F'''(R)\Big) .
\end{aligned} \end{equation}

In the case of  action $S'$ we have
\begin{equation} \begin{aligned}
\overline \rho &= \frac{1}{8 \pi G} \Big( \frac \alpha2 R^2 +  \Lambda - 6 \alpha \frac{\dot a}{a} \dot R \Big) , \\
\overline p &= \frac{-1}{8 \pi G} \Big( \frac \alpha2 R^2 +  \Lambda + 2\alpha (3 \frac{\dot a}{a} \dot R + \ddot R) \Big) , \\
\overline w &= \frac{\overline p}{\overline \rho} = -1 -
\frac{2\alpha (6 \frac{\dot a}{a} \dot R +\ddot R)}{\frac \alpha2
R^2 +  \Lambda - 6 \alpha \frac{\dot a}{a} \dot R} \, .
\end{aligned} \end{equation}

It seems to be natural that any $F(R)$ theory should  satisfy \cite{faraoni}
\begin{align}
F'(R) > 0 , \qquad
-1.1 \leq \bar w \leq -0.98  . \label{cons:2}
\end{align}

In the \textit{Case} 1, we have  $\FF\left(  2 \lambda^2 \right) = 0$ and consequently $\alpha = 0$ and therefore  $ F' (R) = \beta = \frac{1}{2}, \, \,\bar w = -1.$

In the \textit{Case} 2, we have a solution of the form $a(t) = a_0(\sigma
e^{\lambda t} + \tau e^{- \lambda t}) ,$ where the constants $a_0$,
$\sigma$, $\tau$, $\lambda$ are subject to the  constraints
\begin{equation}
\lambda^2 = \frac{1}{3}  \Lambda,  \quad
3 k = 4 a_0^2 \Lambda \sigma \tau .
 \end{equation}
From this set of parameters  we see that
\begin{equation}
R(t) = 4 \Lambda , \qquad
\overline{w} = -1 .
\end{equation}

In the \textit{Case} 3,  we obtain
\begin{equation}
\alpha = \frac{1}{12\Lambda }+\frac{16\pi G C}{3}f_0 ,  \quad \beta = \frac{2}{3} (1 + 16 \pi G C \Lambda f_0) ,
\end{equation}
and
\begin{equation}
\overline w = -1 + h(k,u, v),
\end{equation}
where
\begin{align}
h(k,u,v) &= \frac{512 k u \left(k^2+12 k u+144 u^2\right) \left(v + 1 \right)}{(k^4 + 20736 u^4)V_1 + (48 k^3 u + 6912 k u^3)V_2 + 288 k^2 u^2 V_3 } , \\
V_1 & = 2v +5, \quad V_2 =  14 v +11, \quad V_3 = 70 v + 79  ,\\
u &= \frac{1}{3} a_0^2 \Lambda  \sigma^2 e^{2 \sqrt{\frac{\Lambda}{3}} t  }, \quad
v = 64 \pi G C f_0 \Lambda .
\end{align}

Obviously, in the case $k=0$ we have $\overline{w} = -1$.
Note that $h(-1,u , v) = h(1, -u, v).$ Hence it is sufficient to analyze only $h(1, u, v)$ and it will be presented elsewhere with
some other properties of this $F(R)$ theory.

\section{Discussion and Concluding Remarks}
In this paper we have considered a nonlocal gravity model with cosmological constant $\Lambda$ and without matter. We found three types of
nonsingular bouncing solutions for cosmological scale factor in the form $a(t) = a_0  (\sigma  e^{\lambda t} + \tau e^{-\lambda t} )
.$ Solutions exist for all three values of spatial curvature constant $k = 0, \pm 1 .$ All these solutions depend on cosmological constant
$\Lambda$, which is here an arbitrary positive parameter.

Note that trace equation \eqref{trace:2} can be rewritten in the form
\begin{equation}
A_{1}R + A_{2}(4 \lambda^{2}R^{2}- \dot{R}^{2})+ A_{3}=0 ,
\end{equation}
where
\begin{align}
A_{1}&= - \frac{1}{8 \pi G C}+ 12 \lambda^{2}(3 \mathcal{F}(2
\lambda^{2})-2f_{0})- 96 \lambda^{4}\mathcal{F'}(2 \lambda^{2}) , \\
A_{2}&=\mathcal{F'}(2 \lambda^{2}) ,\\
A_{3}&= \frac{ \Lambda}{2 \pi G C}-144 \lambda^{4}(3\mathcal{F}(2
\lambda^{2})-2f_{0})+ 576 \lambda^{6}\mathcal{F'}(2 \lambda^{2}) .
\end{align}
From $ A_{1}= A_{2}= A_{3}=0$ one obtains the following system of equations:
\begin{align}
&12 \lambda^{2}(3 \mathcal{F}(2 \lambda^{2})-2f_{0})=\frac{1}{8 \pi
G C}, \quad  \quad \mathcal{F'}(2 \lambda^{2}) = 0 , \\
&144 \lambda^{4}(3\mathcal{F}(2 \lambda^{2})-2f_{0})= \frac{
\Lambda}{2 \pi G C}.
\end{align}
Finally, we get
\begin{align}
\lambda^{2}= \frac{\Lambda}{3} , \quad
\mathcal{F}(2 \lambda^{2})= \frac{1}{96 \Lambda \pi G C}+
\frac{2}{3}f_{0} , \quad
\mathcal{F'}(2 \lambda^{2})= 0,
\end{align}
and this is related to the corresponding  conditions in \cite{biswas} and to the first two equations of our \textit{Case} 3. However
in \textit{Case} 3 there is additional condition $k = - 4 a_0^2 \Lambda \sigma \tau ,$ which does not permit solution of hyperbolic cosine type
when $k = 0$. In \cite{biswas0,biswas} this problem was solved adding some radiation in 00 equation of motion.

Note that in the above solutions for the scale factor one can write more general expression by replacement $t \to t-t_0 ,$ i.e.
\begin{equation} \label{t0} a(t) = a_0(\sigma e^{\lambda (t-t_0)} + \tau e^{-\lambda (t-t_0)}) .\end{equation}
When $\sigma > 0$ and $\tau > 0$, and  $t_0=\frac1{2\lambda}\ln(\frac{\sigma}{\tau})$ then \eqref{t0} can be rewritten as
 \begin{equation} \label{cosh} a(t) = 2a_0\sqrt{\sigma\tau}\cosh(\lambda t) .\end{equation}
If $\tau < 0 < \sigma$ and  $t_0=\frac1{2\lambda}\ln(-\frac{\sigma}{\tau})$ then the scale factor  \eqref{t0} becomes
\begin{equation} \label{sinh} a(t) = 2a_0\sqrt{-\sigma\tau}\sinh(\lambda t) .\end{equation}

Note that one can construct a solution $a(t)$ for a negative value of $\Lambda .$ In this article we used linear ansatz \eqref{ansatz}, but
some other ans\"atze \cite{DDGR} can be also useful in search of new cosmological solutions. Here we considered a nonlocal gravity model without
matter, however modern cosmology will probably need both -- modified gravity and an exotic matter (see, e.g. zeta strings \cite{dragovich}
and $p$-adic matter \cite{dragovich1}).

\section{Acknowledgements}
This investigation is partially supported by the Serbian Ministry of Education, Science and
Technological Development under project No 174012, and by the ICTP – SEENET-MTP Grant PRJ-09 ``Cosmology and Strings'' within the framework of the SEENET-MTP Network. B.D. is thankful to the organizers
of the 8-th Workshop ``Quantum Field Theory and Hamiltonian Systems'' (University of Craiova, Romania, 19--22 September 2012)
for stimulating scientific environment  and hospitality, and to the UNESCO - Venice Office for a support. B. D. would also like to thank  A. S. Koshelev and S. Yu. Vernov for useful discussions.

\end{document}